# Plasmon-enhanced Brillouin Light Scattering spectroscopy for magnetic systems. I. Theoretical Model

Valeri Lozovski[1,2] and Andrii V. Chumak[3],

[1] Educational Scientific Institute of High Technologies, Taras Shevchenko National University of Kyiv, 4-g Hlushkova Avenue, Kyiv, 03022, Ukraine
[2] The Erwin Schrödinger International Institute for Mathematics and Physics, Boltzmanngasse 9, 1090 Vienna, Austria
[3] Faculty of Physics, University of Vienna, Boltzmanngasse 5 A, 1090 Vienna, Austria

**Abstract**

Brillouin light scattering (BLS) spectroscopy is an effective method for detecting spin waves in magnetic thin films and nanostructures. While it provides extensive insight into the properties of spin waves, BLS spectroscopy is impeded in many practical cases by the limited range of detectable spin wave wavenumbers and its low sensitivity. Here, we present a generalized theoretical model describing plasmon-enhanced BLS spectroscopy. Three types of plasmonic nanoparticles in the shape of an ellipsoid of rotation are considered: a single plasmon resonator, a sandwiched plasmonic structure in which two nanoparticles are separated by a dielectric spacer, and an ensemble of metallic nanoparticles on the surface of a magnetic film. The effective susceptibilities for the plasmonic systems at the surface of the magnetic film are calculated using the electrodynamic Green functions method, and the enhancement coefficient is defined. It is analytically shown that the ratio of the plasmon resonator height to its radius plays the key role in the development of plasmon-enhanced BLS spectroscopy. The developed model serves as a basis for numerical engineering of the optimized plasmon nanoparticle morphology for BLS enhancement.

**Introduction**

Brillouin light scattering (BLS) spectroscopy is a well-established and effective tool for studying the dynamic properties of magnetic systems [1-6]. It is one of the key methodologies in magnonics – a field that focuses on the properties and behavior of spin waves [7-10] and their quanta magnons. The advantages of the BLS technique include its non-destructive nature, high dynamic range, relatively high frequency, spatial and temporal resolution, as well as the ability to perform wavenumber [11, 12] and phase-resolved [13] measurements. Besides the direct detection of magnon transport, BLS allows for the characterization of magnetic materials (e.g., magnetization, exchange stiffness, internal field inhomogeneity, coupling between magnetic elements, etc.) in various magnetic materials and structures down to the nanoscale [1,3-6, 14].

However, BLS spectroscopy is challenged by the typically low amplitude of the scattered signal, since only a small fraction of the incident photons is non-elastically scattered by the magnons. In addition, it often suffers from insufficient spatial resolution, which is limited by the optical fundamental limit and is usually greater than 300 nm, and from the limited maximum spin-wave wavenumber of the order of 20 rad/µm [15] that can be detected by BLS. The wavenumber limitation is particularly critical given the advantages that the high-wavelength exchange spin waves offer for data transport and processing [16, 17]. Nevertheless, significant progress has been made in the field of BLS spectroscopy in recent years. For example, the use of the near-field approach has been exploited to improve the spatial resolution of the BLS spectroscopy using an AFM cantilever with a tip [18]. To access spin waves of larger wave vectors, subdiffraction confinement of the electromagnetic field was achieved in Ref. [19] by placing a metal nanoantenna on the surface of the Yttrium Iron Garnet (YIG) film. Another promising approach is based on the use of optically induced Mie resonances in nanoparticles, which allowed the detection of spin waves with wavenumbers beyond 50 rad/µm [15]. However, such approaches either do not enhance [15] or reduce the strength of the detected photonic signal [18, 19], which could critically limit the measurement times of BLS spectroscopy.

The improvement of the BLS spectroscopy sensitivity can be achieved by implementing magnon-plasmonic systems [20] and by the use of a local-field enhancement effect in nanoplasmonic systems [21] specifically. Plasmon is a quanta of oscillation of the free electron gas density with respect to the fixed positive ions.

The size of the light spot of the focused laser beam on the film surface is about 300 nm. Given that the intensity of the scattered BLS signal is weak, and a satisfactory spatial resolution can be equal to the linear dimension of the spot, it would be desirable to use the entire area of the test light beam to increase the intensity of the scattered signal. One way to address this challenge involves employing a complex multi-particle plasmonic structure in the form of a monolayer of the metal nanoparticles that are capable of exciting collective modes.

Here we present a generic analytical model describing the enhancement of the effective susceptibility of the magnon-plasmon systems and define the BLS spectroscopy enhancement coefficient. First, a single plasmonic nanoparticle in the form of an ellipsoid of rotation on the surface of a magnetic film is considered, followed by the study of the complex structure when two plasmonic nanoparticles are separated by the dielectric spacer. It is shown that the geometrical dimensions of the plasmonic resonator are of crucial importance and the proposed sandwich structure gives additional degrees of freedom to engineer a required plasmonic resonance. Finally, the array of nearly localized plasmonic resonators is described analytically. The developed theoretical model is used in the second part of these studies [22], which presents the numerical studies of plasmon-enhanced BLS spectroscopy.

## 1. Enhancement of Brillouin Light Scattering (BLS) signal

BLS spectroscopy is based on the phenomenon of inelastic scattering of photons off phonons or magnons in the solid-state media under investigation. This interaction can be detected as a small frequency shift [3] between the inelastically and elastically scattered photons on a frequency scale and accounts for the conservation of energy and momentum:

$$\left.\begin{array}{l}\omega_S = \omega_I \pm \Omega \\ \mathbf{q}_S = \mathbf{q}_I \pm \mathbf{k}\end{array}\right\}, \tag{1}$$

where the sign " – " corresponds to photon scattering with magnon creation, and the sign "+" corresponds to photon scattering with magnon annihilation, $\hbar\omega_I$ is the photon's energy of an incident laser light, $\hbar\omega_S$ is the energy of a scattered photon, $\hbar\Omega$ is the magnon's energy, $\mathbf{k}$ is the wave vector of a magnon, $\mathbf{q}_I$ is the photon wave vector of an incident laser light, $\mathbf{q}_S$ is the wave vector of a scattered photon.

The conservation laws imply that the wave vector of the magnon depends on the angle of the incident probe laser light in the backscattering mode ($\mathbf{q}_S = -\mathbf{q}_I$) modus operandi $q = 2q_I \sin\vartheta$, with $q$ in-plane component of the wave vector. If the investigated magnetic material is a thin film, the normal to the film's plane component of the magnon wave vector is undefined and only the in-plane components of the magnon wave vector have a physical sense.

The energy of a magnon is much lower than the energy of a photon, so it can be assumed that both the Stokes – $\hbar\bar{\omega} = \hbar(\omega - \Omega)$ and anti-Stokes – $\hbar\tilde{\omega} = \hbar(\omega + \Omega)$ components of the BLS have approximately the frequency of the probing laser light. In the following, we only consider the Stokes components.

The model of dynamic excitation of the magnetization in the magnetic material induces a small, frequency-dependent change in the susceptibility tensor [2]. In the case of spin-wave scattering, the change in the susceptibility tensor is induced by the magneto-optical coupling. The magnetization-dependent part of the polarization $P_i(\mathbf{R},\bar{\omega})$ induced by the electric field of the incident light $E_j(\mathbf{R},\omega)$ can be written as:

$$P_i(\mathbf{R},\bar{\omega}) = \frac{1}{4\pi}\chi_{ij} E_j(\mathbf{R},\omega), \tag{2}$$

$E_j$ is a component of the electric field inside the film, $\chi_{ij}$ is the susceptibility tensor in the first-order magneto-optical effects and its relation to the Kerr effect is given by [2]:

$$\chi_{ij} = \begin{pmatrix} 0 & KM_3 & -KM_2 \\ -KM_3 & 0 & KM_1 \\ KM_2 & -KM_1 & 0 \end{pmatrix}, \tag{3}$$

where $M_1$, $M_2$ and $M_3$ are the magnetization (components) along the *x*, *y* and *z* directions ($\bar{\omega} = \omega - \Omega$), $K$ is the constant of the Kerr effects.

Thus, the oscillating electric dipole $P_i(\mathbf{R},\bar{\omega})$ in Eq. (2) can be considered as the source of BLS (shown as red arrow in Fig. 1). And using the electrodynamic Green function method one can connect the dipole $P_i(\mathbf{R},\bar{\omega})$ and the field of BLS (the connection is provided further in Eq.(22)).

The BLS signal typically has a very low intensity and is difficult to measure, therefore signal enhancement poses an urgent challenge in the field of magneto-optics. One way to achieve this enhancement is to use plasmonic resonance in nanostructures.

Consider a standard BLS experiment in which the measured signal is induced by a photon-magnon interaction (see Fig. 1). The plasmonic system is located at the surface of a magnetic film. The enhancement of the BLS signal can be provided under the condition of plasmonic resonance. Therefore, the effective susceptibilities (linear responses to the external relatively the particle field) of the plasmonic systems should be calculated to analyze the resonance conditions. In the following, we will use the notations: '1' – for GGG substrate characterized in optical range by dielectric constant $\varepsilon_1$, '2' – for YIG thin film characterized in optical range by dielectric constant $\varepsilon_2$, and '3' – for the environment characterized (air) by dielectric constant $\varepsilon_3$ (Fig.1).

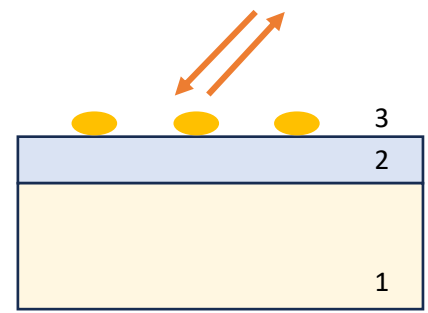

Fig. 1. Sketch of the measured BLS signal in a backscattering geometry.

There are two approaches to achieve the resonance conditions:

i. If the magnetic system consists of 1) a thin stripe of width about a wavelength of a magnon [23] together with 2) a nanoplasmonic structure made of a few nanoparticles interacting weakly with one another and situated at the surface of magnet film, then the plasmonic resonance is defined by a single plasmonic structure's morphology.

ii. If the magnetic system is composed of a submonolayer of nanoparticles on top of the magnetic film surface, which one can suppose as infinitive in the surface plane, then the resonance will have the character of a configurational resonance [24] and the conditions for its formation will depend on the structure of a single nanoobject and the particles concentration.

## 2. Effective susceptibility of plasmonic systems

Two kinds of plasmonic structures will be considered in this work. Thus, two solutions to the BLS signal enhancement challenge should be provided according to the methodology provided in Chapter 1.

### 2.1. Single plasmonic structure

There are cases when only a limited number of nanostructures can be placed on the surface of a magnetic film, for example, if the magnetic film is a very thin, narrow strip (see, e.g., [14]). In this case, it can be assumed that the system consists of one plasmon structure on the surface of the magnetic strip. It is obvious that then the resonant enhancing of the BLS signal will depend mainly on the composition of the nanostructure (Fig. 2). The nanostructure at the surface can be either a single nanoparticle or a complex nanoparticle system, consisting of several elements.

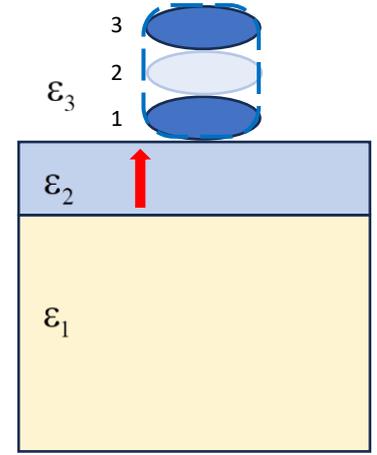

Fig. 2. Sketch of a single nanostructure at the magnetic film. Arrow means the oscillating electric dipole moment Eq. (2).

First, let us consider a single nanoparticle at the surface of a magnetic film. Let $\chi(\omega)$ be the dielectric response of the material of which the nanoparticle is fabricated. According to the approach, developed in Ref. [25], the effective susceptibility of the nanoparticle can be calculated with the expression:

$$X_{ij}^{(S)}(\omega, \mathbf{R}) = \left[ \chi^{-1}(\omega)\delta_{ji} - \int_{V_p} d\mathbf{R}' G_{ji}^{(33)}(\mathbf{R}, \mathbf{R}', \omega) \right]^{-1}, \qquad (4)$$

where $G_{ij}^{(33)}(\mathbf{R}, \mathbf{R}', \omega)$ is the electrodynamic Green function of the film at the substrate, if the point source and the observation points are located in the environment (medium '3'), characterized by the dielectric constant $\varepsilon_3$; integration is over the volume of nanoparticle $V_p$. Under the condition of plasmonic resonance:

$$\operatorname{Re}\det\left[\chi^{-1}(\omega)\delta_{ji} - \int_V d\mathbf{R}' G^{(33)}_{ji}(\mathbf{R},\mathbf{R}',\omega)\right] = 0 \ , \tag{5}$$

where $\mathbf{R} \in V_p$, the amplitude of the scattered field can be enhanced. From this perspective, Eq. (5) can be regarded as a necessary condition for enhancing the BLS response. It is evident, that Eq. (5) can be used as a starting point of the optimization, as the plasmonic resonance condition depends also on the shape, dimension and material of the nanoparticle. This problem will be solved numerically in the 2$^{nd}$ part of the work.

Consider the nanostructure made of three oblate ellipsoids that are located one above the other. It should be noted that we use an ellipsoid as an object in order to obtain an analytical solution to the problem. In practice, the results obtained can be qualitatively projected onto the disks, which can be easily fabricated using modern nanostructuring techniques. However, the precise qualitative analysis for the disks would require specialized numerical studies. Using the methodology developed in Ref. [25], one can find the effective susceptibility of the structure. Let the external field $E_i^{(0)}(\mathbf{R})$ act on the system. The local field in arbitrary point of the considered system can be found from the equation of self-consistency [26]:

$$\begin{aligned}
E_i(\mathbf{R},\omega) = E_i^{(0)}(\mathbf{R},\omega) & - \\
& - k_0^2 \int_{V_1} d\mathbf{R}' G^{(33)}_{ij}(\mathbf{R},\mathbf{R}',\omega)\chi^{(1)}_{jl}(\omega)E_l(\mathbf{R}',\omega) - \\
& - k_0^2 \int_{V_2} d\mathbf{R}' G^{(33)}_{ij}(\mathbf{R},\mathbf{R}',\omega)\chi^{(2)}_{jl}(\omega)E_l(\mathbf{R}',\omega) - \\
& - k_0^2 \int_{V_3} d\mathbf{R}' G^{(33)}_{ij}(\mathbf{R},\mathbf{R}',\omega)\chi^{(3)}_{jl}(\omega)E_l(\mathbf{R}',\omega) \ ,
\end{aligned} \tag{6}$$

where $\chi^{(\alpha)}_{ij}(\omega)$, $\alpha = 1,2,3$ are the susceptibilities of single nanoparticles forming the plasmonic structure (Fig. 2), $k_0 = \omega/c$ with $c$ light velocity. Integrations in Eq. (6) are over the volumes of nanoparticles 1, 2 and 3. Let $X^{(S)}_{ij}(\mathbf{R},\omega)$ be the effective susceptibility of the plasmonic structure, then the electric current induced in each of the nanoparticles can be written as:

$$J_i^{(\alpha)}(\mathbf{R},\omega) = -i\omega X_{ij}^{(S)}(\mathbf{R},\omega)E_j^{(0)}(\mathbf{R},\omega) = -i\omega \chi_{ij}^{(\alpha)}(\omega)E_j(\mathbf{R},\omega) \ . \tag{7}$$

Thus, one can establish the connection between the field, external relative to the plasmonic structure $E_j^{(0)}(\mathbf{R},\omega)$ and the local field $E_i(\mathbf{R},\omega)$:

$$E_i^{(0)}(\mathbf{R}^{(\alpha)},\omega) = \left[X_{ji}^{(S)}(\mathbf{R}^{(\alpha)},\omega)\right]^{-1}\chi_{jl}^{(\alpha)}(\omega)E_l(\mathbf{R}^{(\alpha)},\omega) \ . \tag{8}$$

Further, by substituting this expression in Eq. (6) and assuming $\mathbf{R}=\mathbf{R}^{(1)}$, one can obtain a self-consisting field inside the particle '1':

$$\begin{aligned}
E_i(\mathbf{R}^{(1)},\omega) = & \left[X_{ji}^{(S)}(\mathbf{R}^{(1)},\omega)\right]^{-1}\chi_{jl}^{(1)}(\omega)E_l(\mathbf{R}^{(1)},\omega) - \\
& - k_0^2 \int_{V_1} d\mathbf{R}' G_{ij}^{(33)}(\mathbf{R}^{(1)},\mathbf{R}',\omega)\chi_{jl}^{(1)}(\omega)E_l(\mathbf{R}',\omega) - \\
& - k_0^2 \int_{V_2} d\mathbf{R}' G_{ij}^{(33)}(\mathbf{R}^{(1)},\mathbf{R}',\omega)\chi_{jl}^{(2)}(\omega)E_l(\mathbf{R}',\omega) - \\
& - k_0^2 \int_{V_3} d\mathbf{R}' G_{ij}^{(33)}(\mathbf{R}^{(1)},\mathbf{R}',\omega)\chi_{jl}^{(3)}(\omega)E_l(\mathbf{R}',\omega).
\end{aligned} \tag{9}$$

Note, that the last two terms in this equation are the fields induced inside the first nanoparticle by the currents inside the nanoparticles '2' and '3'. Similarly, one can formulate the equations describing the fields inside particles '2' and '3':

$$\begin{aligned}
E_i(\mathbf{R}^{(2)},\omega) = & \left[X_{ji}^{(S)}(\mathbf{R}^{(2)},\omega)\right]^{-1}\chi_{jl}^{(2)}(\omega)E_l(\mathbf{R}^{(2)},\omega) - \\
& - k_0^2 \int_{V_1} d\mathbf{R}' G_{ij}^{(33)}(\mathbf{R}^{(2)},\mathbf{R}',\omega)\chi_{jl}^{(1)}(\omega)E_l(\mathbf{R}',\omega) - \\
& - k_0^2 \int_{V_2} d\mathbf{R}' G_{ij}^{(33)}(\mathbf{R}^{(2)},\mathbf{R}',\omega)\chi_{jl}^{(2)}(\omega)E_l(\mathbf{R}',\omega) - \\
& - k_0^2 \int_{V_3} d\mathbf{R}' G_{ij}^{(33)}(\mathbf{R}^{(2)},\mathbf{R}',\omega)\chi_{jl}^{(3)}(\omega)E_l(\mathbf{R}',\omega) \ ,
\end{aligned} \tag{10}$$

$$\begin{aligned}
E_i(\mathbf{R}^{(3)},\omega) = & \left[X_{ji}^{(S)}(\mathbf{R}^{(3)},\omega)\right]^{-1}\chi_{jl}^{(3)}(\omega)E_l(\mathbf{R}^{(3)},\omega) - \\
& - k_0^2 \int_{V_1} d\mathbf{R}' G_{ij}^{(33)}(\mathbf{R}^{(3)},\mathbf{R}',\omega)\chi_{jl}^{(1)}(\omega)E_l(\mathbf{R}',\omega) - \\
& - k_0^2 \int_{V_2} d\mathbf{R}' G_{ij}^{(33)}(\mathbf{R}^{(3)},\mathbf{R}',\omega)\chi_{jl}^{(2)}(\omega)E_l(\mathbf{R}',\omega) - \\
& - k_0^2 \int_{V_3} d\mathbf{R}' G_{ij}^{(33)}(\mathbf{R}^{(3)},\mathbf{R}',\omega)\chi_{jl}^{(3)}(\omega)E_l(\mathbf{R}',\omega).
\end{aligned} \tag{11}$$

Then, if to integrate the Eq. (9) over the volume of the first particle, the Eq. (10) over the volume of the second particle and the Eq. (11) over the volume of the third particle, one can obtain:

$$\int_{V_1} d\mathbf{R}^{(1)} E_i(\mathbf{R}^{(1)},\omega) = \int_{V_1} d\mathbf{R}^{(1)} \left[ X_{ji}^{(S)}(\mathbf{R}^{(1)},\omega) \right]^{-1} \chi_{jl}^{(1)}(\omega) E_l(\mathbf{R}^{(1)},\omega) -$$
$$- k_0^2 \int_{V_1} d\mathbf{R}^{(1)} \int_{V_1} d\mathbf{R}' G_{ij}^{(33)}(\mathbf{R}^{(1)},\mathbf{R}',\omega) \chi_{jl}^{(1)}(\omega) E_l(\mathbf{R}',\omega) -$$
$$- k_0^2 \int_{V_1} d\mathbf{R}^{(1)} \int_{V_2} d\mathbf{R}' G_{ij}^{(33)}(\mathbf{R}^{(1)},\mathbf{R}',\omega) \chi_{jl}^{(2)}(\omega) E_l(\mathbf{R}',\omega) -$$
$$- k_0^2 \int_{V_1} d\mathbf{R}^{(1)} \int_{V_3} d\mathbf{R}' G_{ij}^{(33)}(\mathbf{R}^{(1)},\mathbf{R}',\omega) \chi_{jl}^{(3)}(\omega) E_l(\mathbf{R}',\omega). \quad (12)$$

$$\int_{V_2} d\mathbf{R}^{(2)} E_i(\mathbf{R}^{(2)},\omega) = \int_{V_2} d\mathbf{R}^{(2)} \left[ X_{ji}^{(S)}(\mathbf{R}^{(2)},\omega) \right]^{-1} \chi_{jl}^{(2)}(\omega) E_l(\mathbf{R}^{(2)},\omega) -$$
$$- k_0^2 \int_{V_2} d\mathbf{R}^{(2)} \int_{V_1} d\mathbf{R}' G_{ij}^{(33)}(\mathbf{R}^{(2)},\mathbf{R}',\omega) \chi_{jl}^{(1)}(\omega) E_l(\mathbf{R}',\omega) -$$
$$- k_0^2 \int_{V_2} d\mathbf{R}^{(2)} \int_{V_2} d\mathbf{R}' G_{ij}^{(33)}(\mathbf{R}^{(2)},\mathbf{R}',\omega) \chi_{jl}^{(2)}(\omega) E_l(\mathbf{R}',\omega) -$$
$$- k_0^2 \int_{V_2} d\mathbf{R}^{(2)} \int_{V_3} d\mathbf{R}' G_{ij}^{(33)}(\mathbf{R}^{(2)},\mathbf{R}',\omega) \chi_{jl}^{(3)}(\omega) E_l(\mathbf{R}',\omega). \quad (13)$$

$$\int_{V_3} d\mathbf{R}^{(3)} E_i(\mathbf{R}^{(3)},\omega) = \int_{V_3} d\mathbf{R}^{(3)} \left[ X_{ji}^{(S)}(\mathbf{R}^{(3)},\omega) \right]^{-1} \chi_{jl}^{(3)}(\omega) E_l(\mathbf{R}^{(3)},\omega) -$$
$$- k_0^2 \int_{V_3} d\mathbf{R}^{(3)} \int_{V_1} d\mathbf{R}' G_{ij}^{(33)}(\mathbf{R}^{(3)},\mathbf{R}',\omega) \chi_{jl}^{(1)}(\omega) E_l(\mathbf{R}',\omega) -$$
$$- k_0^2 \int_{V_3} d\mathbf{R}^{(3)} \int_{V_2} d\mathbf{R}' G_{ij}^{(33)}(\mathbf{R}^{(3)},\mathbf{R}',\omega) \chi_{jl}^{(2)}(\omega) E_l(\mathbf{R}',\omega) -$$
$$- k_0^2 \int_{V_3} d\mathbf{R}^{(3)} \int_{V_3} d\mathbf{R}' G_{ij}^{(33)}(\mathbf{R}^{(3)},\mathbf{R}',\omega) \chi_{jl}^{(3)}(\omega) E_l(\mathbf{R}',\omega). \quad (14)$$

Consider the term in the right part of Eq.(12), indicated by an underlining with a single line – (…), and similar term (…) in the right part of Eq. (13). The term (…) in Eq. (12) is the average field at the particle '1' which is induced by the currents in the particle '2'. Simultaneously, the term (…) in Eq. (13) is the average field in the particle '2' induced by the currents in the particle '1'. According to the reciprocal theorem [27], one can expect these fields to be equal one to another. Thus, we can substitute a field (...) from Eq. (13) into a similar term in Eq. (12). Similarly, we can operate with the terms indicated by underlining with a double line (...) in Eq. (12) and Eq. (14). As a result, we arrive to:

$$\int_{V_1} d\mathbf{R}^{(1)} \left\{ \left[ X_{ji}^{(S)}(\mathbf{R}^{(1)},\omega) \right]^{-1} \chi_{jl}^{(1)}(\omega) - \delta_{il} - \right.$$
$$- k_0^2 \int_{V_1} d\mathbf{R}' G_{ij}^{(33)}(\mathbf{R}',\mathbf{R}^{(1)},\omega)\chi_{jl}^{(1)}(\omega) -$$
$$- k_0^2 \int_{V_2} d\mathbf{R}^{(2)} G_{ij}^{(33)}(\mathbf{R}^{(2)},\mathbf{R}^{(1)},\omega)\chi_{jl}^{(1)}(\omega) -$$
$$\left. - k_0^2 \int_{V_3} d\mathbf{R}^{(3)} G_{ij}^{(33)}(\mathbf{R}^{(3)},\mathbf{R}^{(1)},\omega)\chi_{jl}^{(1)}(\omega) \right\} E_l(\mathbf{R}^{(1)},\omega) = 0. \tag{15}$$

The local field $E_l(\mathbf{R}^{(1)},\omega)$ can be presented as a Fourier set $E_l(\mathbf{R}^{(1)},\omega) = \sum_{\mathbf{k}} E_l^{(\mathbf{k})} e^{-i\mathbf{k}\mathbf{R}^{(1)}}$ and then substituted in the Eq. (15), so that:

$$\int_{V_1} d\mathbf{R}^{(1)} \sum_{\mathbf{k}} \left\{ \left[ X_{ji}^{(S)}(\mathbf{R}^{(1)},\omega) \right]^{-1} \chi_{jl}^{(1)}(\omega) - \delta_{il} - \right.$$
$$- k_0^2 \int_{V_1} d\mathbf{R}' G_{ij}^{(33)}(\mathbf{R}',\mathbf{R}^{(1)},\omega)\chi_{jl}^{(1)}(\omega) -$$
$$- k_0^2 \int_{V_2} d\mathbf{R}^{(2)} G_{ij}^{(33)}(\mathbf{R}^{(2)},\mathbf{R}^{(1)},\omega)\chi_{jl}^{(1)}(\omega) -$$
$$\left. - k_0^2 \int_{V_3} d\mathbf{R}^{(3)} G_{ij}^{(33)}(\mathbf{R}^{(3)},\mathbf{R}^{(1)},\omega)\chi_{jl}^{(1)}(\omega) \right\} E_l^{(\mathbf{k})} e^{-i\mathbf{k}\mathbf{R}^{(1)}} = 0. \tag{16}$$

Because the exponents are the complete set of the orthogonal functions, to satisfy this equation, the following equation should be considered:

$$\left[ X_{ji}^{(S)}(\mathbf{R}^{(1)},\omega) \right]^{-1} \chi_{jl}^{(1)}(\omega) = \delta_{il} +$$
$$+ k_0^2 \int_{V_1} d\mathbf{R}' G_{ij}^{(33)}(\mathbf{R}',\mathbf{R}^{(1)},\omega)\chi_{jl}^{(1)}(\omega) +$$
$$+ k_0^2 \int_{V_2} d\mathbf{R}^{(2)} G_{ij}^{(33)}(\mathbf{R}^{(2)},\mathbf{R}^{(1)},\omega)\chi_{jl}^{(1)}(\omega) +$$
$$+ k_0^2 \int_{V_3} d\mathbf{R}^{(33)} G_{ij}^{(B)}(\mathbf{R}^{(3)},\mathbf{R}^{(1)},\omega)\chi_{jl}^{(1)}(\omega) \tag{17}$$

Consequently, one obtains the effective susceptibility of the plasmonic structure at the surface of a magnetic film $X_{ji}^{(S)}(\mathbf{R},\omega)$ when $\mathbf{R}=\mathbf{R}^{(1)}$:

$$X_{ji}^{(S)}(\mathbf{R}^{(1)},\omega) = \left\{ \left[ \chi_{ji}^{(1)}(\omega) \right]^{-1} + k_0^2 \int_{V_1} d\mathbf{R}' G_{ij}^{(33)}(\mathbf{R}',\mathbf{R}^{(1)},\omega) + \right. $$
$$\left. + k_0^2 \int_{V_2} d\mathbf{R}^{(2)} G_{ij}^{(33)}(\mathbf{R}^{(2)},\mathbf{R}^{(1)},\omega) + k_0^2 \int_{V_3} d\mathbf{R}^{(3)} G_{ij}^{(33)}(\mathbf{R}^{(3)},\mathbf{R}^{(1)},\omega) \right\}^{-1} \quad (18)$$

Analogously the effective susceptibility is calculated regarding the system $X_{ji}^{(S)}(\mathbf{R},\omega)$ for $\mathbf{R}=\mathbf{R}^{(2)}$ and $\mathbf{R}=\mathbf{R}^{(3)}$

$$X_{ji}^{(S)}(\mathbf{R}^{(\alpha)},\omega) = \left\{ \left[ \chi_{ji}^{(\alpha)}(\omega) \right]^{-1} + k_0^2 \int_{V_\alpha} d\mathbf{R}' G_{ij}^{(33)}(\mathbf{R}',\mathbf{R}^{(\alpha)},\omega) + \right. $$
$$\left. + k_0^2 \int_{V_1} d\mathbf{R}^{(1)} G_{ij}^{(33)}(\mathbf{R}^{(1)},\mathbf{R}^{(\alpha)},\omega) + k_0^2 \int_{V_\beta} d\mathbf{R}^{(\beta)} G_{ij}^{(33)}(\mathbf{R}^{(\beta)},\mathbf{R}^{(\alpha)},\omega) \right\}^{-1}, \quad \alpha \neq \beta, \; \alpha,\beta = 2,3. \quad (19)$$

It should be highlighted that the expressions for the effective susceptibility of the complex plasmon structure contain all possible electrodynamic interactions between the parts of the structure.

## 2.2. Plasmonic structure consisting of the sub-monolayer surface cover by metal nanoparticles

If the plasmonic system is a sub-monolayer surface cover by metal nanoparticles, then these surface particles, shaped as the ellipsoids of rotation, have effective susceptibility $\tilde{\chi}_{ij}$ [25]. According to Ref. [24], the effective susceptibility of a submonolayer cover of the nanoparticles, can be calculated as:

$$X_{ij}(\overline{\omega},k,d) = \left[ \left( \tilde{\chi}_{ii_{zz}}(\omega) \right)^{-1} \delta_{ji_{zz}} + nk_0^2 G_{ji}^{(33)}(k,d,\overline{\omega}) \right]^{-1}, \quad (20)$$

where $n$ is the concentration of nanoparticles at the surface of the magnetic film, $d$ is the distance from the surface of the film and center of nanoparticle, and $G_{ji}^{(33)}(k,d,\overline{\omega})$ is the electromagnetic Green function of the system 'film at the substrate' which was calculated in Ref. [28] when wave vector *k* is directed along OX axis lying in the plane of the film.

In all assumptions above, the BLS signal is acquired assuming under a backscattering condition. Wave vector 2*k* in Eq. (20) is the wave vector of a magnon excited by a photon of a probing light. In addition, the surface wave (e.g., plasmons in the metal surface or polaritons in dielectrics including magnetic surface) can be excited by the direct laser irradiation of the surface covered by the metal nanoparticles [29].

## 3. Enhancement of BLS signal with plasmonic system under the 'plasmonic resonance' conditions

In this chapter, we consider the enhancement of BLS with the plasmonic system under the 'plasmonic resonance' condition. Herein, the term 'plasmonic resonance' should be understood as the state of the plasmonic system when the scattered (BLS) field reaches a maximum. The source of the scattered field according to Ref. [2] is the oscillating electric dipole, caused by a magnon-photon interaction. The value of this dipole moment was established in Ref. [2] and it is connected to the electric field with Eq. (2). Therefore, the BLS field can be written via the electrodynamic Green function $G_{ij}^{(32)}(\mathbf{R}, \mathbf{R}', \bar{\omega})$ of the film at the substrate:

$$E_i^{(BLS-0)}(\mathbf{R}, \bar{\omega}) = \int_{V_{\text{int}}} d\mathbf{R}' G_{ij}^{(32)}(\mathbf{R}, \mathbf{R}', \bar{\omega}) P_j(\mathbf{R}', \bar{\omega}) \ , \qquad (21)$$

where the superscript '(32)' means that the source of the field is located inside the film (medium '2') and the observation point is located in the environment (medium '3'). The integration is over the 'interaction volume' – region inside the magnetic film, in which the photon-magnon scattering processes occur. When the plasmonic structure is located at the film surface, the field Eq. (21) will be an external field acting on the structure. Then, the effective dipole moment induced inside the plasmonic system, based on the field in Eq. (22), can be considered as a secondary field source:

$$P_j^{pl}(\mathbf{R}, \bar{\omega}) = \varepsilon_0 X_{jl}^{(S)}(\mathbf{R}, \bar{\omega}) E_l^{(BLS-0)}(\mathbf{R}, \bar{\omega}). \qquad (22)$$

Supposing that the enhanced BLS signal is induced by a dipole momentum Eq. (22), the electric field of BLS signal is calculated according to ($\mu_0 \varepsilon_0 = 1/c^2$ $\bar{k}_0 = \bar{\omega}/c$):

$$E_i^{(sc-plasmon)}(\mathbf{R},\bar{\omega}) = -\bar{\omega}^2 \mu_0 \int_{V_{pl}} d\mathbf{R}' G_{ij}^{(33)}(\mathbf{R},\mathbf{R}',\bar{\omega}) P_j^{(pl)}(\mathbf{R}',\bar{\omega}) =$$

$$= -\bar{k}_0^2 \int_{V_{pl}} d\mathbf{R}' G_{ij}^{(33)}(\mathbf{R},\mathbf{R}',\bar{\omega}) X_{jl}^{(s)}(\mathbf{R}',\bar{\omega}) E_l^{(BLS-0)}(\mathbf{R}',\bar{\omega}) .$$

(23)

Let us look closely at the electrical field $E$ in Eq. (2). This field causes the formation of the oscillating dipole momentum and consists of two parts – the 'external' field and the field enhanced by the plasmon structure field:

$$E_i^{(incom)}(\mathbf{R},\omega) = E_i^{(0)}(\mathbf{R},\omega) - k_0^2 \int_{V_p} d\mathbf{R}' G_{ij}^{(23)}(\mathbf{R},\mathbf{R}',\omega) X_{jl}^{(p)}(\mathbf{R}',\omega) E_l^{(0)}(\mathbf{R}',\omega), \quad (24)$$

where $G_{ij}^{(23)}(\mathbf{R},\mathbf{R}',\omega)$ is the electrodynamic Green function of the film deposited at the substrate, in which the source of the field is located in the medium ('3') and the observation point – in the film ('2'). Integration is over the volume of the plasmon structure. Thus, the dipole moment formed by a photon-magnon scattering is:

$$P_i(\mathbf{R},\bar{\omega}) = \frac{1}{4\pi}\chi_{ij} E_j^{(0)}(\mathbf{R},\omega) - \frac{k_0^2}{4\pi}\chi_{ik} \int_{V_p} d\mathbf{R}' G_{kj}^{(23)}(\mathbf{R},\mathbf{R}',\omega) X_{jl}^{(p)}(\mathbf{R}',\omega) E_l^{(0)}(\mathbf{R}',\omega) . \quad (25)$$

Substituting this expression in Eq. (23), one obtains:

$$E_i^{(sc-plasmon)}(\mathbf{R},\bar{\omega}) = -\frac{\bar{k}_0^2}{4\pi}\int_{V_{pl}} d\mathbf{R}' G_{ij}^{(33)}(\mathbf{R},\mathbf{R}',\bar{\omega}) X_{jk}^{(pl)}(\mathbf{R}',\bar{\omega}) \chi_{kl} E_l^{(0)}(\mathbf{R}',\omega) -$$

$$+\frac{\bar{k}_0^2}{4\pi}\int_{V_{pl}} d\mathbf{R}' G_{ij}^{(33)}(\mathbf{R},\mathbf{R}',\bar{\omega}) X_{jk}^{(pl)}(\mathbf{R}',\bar{\omega}) k_0^2 \int_{V_{int}} d\mathbf{R}'' G_{km}^{(32)}(\mathbf{R}',\mathbf{R}'',\bar{\omega}) \chi_{mn} \times \quad (26)$$

$$\times \int_{V_{pl}} d\mathbf{R}''' G_{nq}^{(23)}(\mathbf{R}'',\mathbf{R}''',\omega) X_{ql}^{(pl)}(\mathbf{R},\omega) E_l^{(0)}(\mathbf{R}',\omega) .$$

From Eq. (26) it is evident that under the condition of plasmon resonance, when $X_{jl}^{(s)}(\mathbf{R}',\omega)$ takes large values, the scattered field can be enhanced. Moreover, enhancement will occur in both terms of Eq. (26).

Ergo, reaching conditions of plasmon resonance can significantly improve the quality of the BLS signal. Thus, determining the conditions for achieving the maximum value of the effective susceptibility of the plasmon system (achieving the minimum of its denominator) can be considered a necessary condition for the enhancement of the BLS signal.

As discussed above, the interaction between the photons and magnons leads to the formation of the oscillating electric dipole inside the film. Then, the electric field induced by the dipole moment can be calculated with Eq. (21). If the plasmonic structure is situated at the surface of the film, then the electric field of the BLS signal consists of two parts – the field as described in Eq. (26) and the field from Eq. (23) induced by plasmonic structure:

$$E_i^{sc-total}(\mathbf{R},\bar{\omega}) = E_i^{sc}(\mathbf{R},\bar{\omega}) + E_i^{sc-plasmon}(\mathbf{R},\bar{\omega}) \ . \tag{27}$$

Then, the amplitude enhancement coefficient is described as:

$$K_{ampl} = \frac{E_i^{sc}(\mathbf{R},\bar{\omega}) + E_i^{sc-plasmon}(\mathbf{R},\bar{\omega})}{E_i^{sc}(\mathbf{R},\bar{\omega})} \ . \tag{28}$$

Likewise, one can define the enhancement coefficient over intensity:

$$K = \frac{\left|E_i^{sc}(\mathbf{R},\omega) + E_i^{sc-plasmon}(\mathbf{R},\omega)\right|^2}{\left|E_i^{sc}(\mathbf{R},\omega)\right|^2} \ . \tag{29}$$

Following these equations, it is apparent that in any case the BLS field can be enhanced under conditions of high values of effective susceptibility $X_{jl}^{(pl)}(\mathbf{R},\omega)$. To satisfy these conditions, the pole part of effective susceptibility should be minimal. Concerning Eq. (26), the 'plasmonic resonance' plays a role at the fundamental frequency $\omega$ and at the shifted frequency $\bar{\omega}$. Because these frequencies are approximately (with an accuracy of about $10^{-5}$) equal one to another, we will further consider a fundamental frequency case, which formally describes the conditions of plasmon resonance. To determine the condition of the optimal parameters of the nanoplasmonic structure for effective enhancing of the BLS, it is necessary to introduce an objective function. For complex nanostructure, for example, consisting of three parts we can use as the objective function (obviously, when the nanostructure consists of a single metal nanoparticle one must take one term in the sum)

$$F_{compl}^{(r)}(\omega) = \sum_{1}^{3} \left| \operatorname{Re} D^{(\alpha)}(R^{(\alpha)},\omega) \right|, \qquad (30)$$

or

$$F_{compl}(\omega) = \sum_{a=1}^{3} \left\{ \left| D^{(a)}(\mathbf{R}^{(a)},\omega) \right| \right\}, \qquad (31)$$

with

$$\det\left\{ \left[ \chi_{ji}^{(a)}(\omega) \right]^{-1} + k_0^2 \int_{V_a} d\mathbf{R}' G_{ij}^{(B)}(\mathbf{R}',\mathbf{R}^{(a)},\omega) + k_0^2 \sum_{b=1,b\neq a}^{3} \int_{V_2} d\mathbf{R}^{(b)} G_{ij}^{(B)}(\mathbf{R}^{(b)},\mathbf{R}^{(a)},\omega) \right\} =$$
$$= D^{(a)}(\mathbf{R}^{(a)},\omega), \quad a=1,2,3. \qquad (32)$$

Thus, the problem of optimization will be reduced to finding the parameters of the plasmon structure at which the function $F(\omega)$ reaches a minimum. In the discussed problem the parameters of nano-structure (at fixed wavelength of testing light) can be materials, dimensions and shapes of the parts of the structure.

The condition formulated as a minimum of objective functions from Eqs. (30)-(31) allows to obtain higher enhancement because it indicates a higher value of effective susceptibility of the nanoplasmonic structure under consideration. The pole part of effective susceptibility $D^{(a)}(\mathbf{R}^{(a)},\omega)$ consists of two types of mass operators, with the first one:

$$M_{ij}^{(sa\_\alpha-\alpha)}(\mathbf{R}^{(\alpha)},\omega) = k_0^2 \int_{V_\alpha} d\mathbf{R}' G_{ij}^{(B)}(\mathbf{R}',\mathbf{R}^{(\alpha)},\omega), \; \alpha=1,2,3 \qquad (33)$$

describing the self-acting processes, i.e., the field induced by the currents inside one particle which acts on the same particle, and the second operator:

$$M_{ij}^{(\text{int}\_\alpha-\beta)}(\mathbf{R}^{(\beta)},\omega) = k_0^2 \int_{V_\alpha} d\mathbf{R}^{(\alpha)} G_{ij}^{(B)}(\mathbf{R}^{(\alpha)},\mathbf{R}^{(\beta)},\omega), \quad \alpha\neq\beta, \; \alpha,\beta=1,2,3 \qquad (34)$$

describing the interactions between the different particles of plasmonic structure. Note that minimum of $F(\omega)$ is not the conditions for 'pure' local plasmon resonances in nanoparticles forming a plasmon nanostructure. The conditions of a local plasmon resonance have a slightly different formal form, namely:

$$\operatorname{Re} D^{(a)}(\mathbf{R}^{(a)}, \omega) = 0, \quad a = 1, 2, 3. \tag{35}$$

So far we have presented three main conditions for determining the morphology of the nanoplasmonic structure (size, shape, and materials of nanoparticles), under which BLS plasmonic enhancement can be realized for a certain wavelength of the probing external laser light. Further improvement of the BLS signal requires to optimize structural morphology, which is discussed in detail below.

As an example, we start the optimization by calculating the dependence of objective function $F_s(\omega)$ of the complex plasmonic structure at the surface of YIG film (with thickness $t_{YIG}$ = 100 nm) on the material, and shape of the nanoparticles from which the nanostructure is consisted of. We fix the radii of nanoparticles as 10 nm and consider two shapes of the spacer oblate ellipsoid - when $h/r$=0.25 and prolate ellipsoid - when $h/r$=2. We suppose that wavelength of testing light is 400 nm. We considered the metal nanoparticles fabricated with gold and silver. The results of calculations of dependence of objective function for the plasmon structure depicted in Fig. 2 on the shape of metal nanoparticles are presented in Fig. 3 and Fig. 4. We can see that both objective functions give similar results. These results confirm our previous reasoning that a small change in the parameters of the nanoplasmonic structure or the replacement of one material by another greatly affects the possibility of BLS signal enhancement. That is, the problem of using a nanoplasmonic structure to enhance the BLS signal should first be considered from the point of view of the fundamental possibility of signal enhancement. After choosing the optimal nanoparticle material, one shall consider to optimize the design of the plasmonic structure [30]. More detailed calculations will be performed in part 2 of this work.

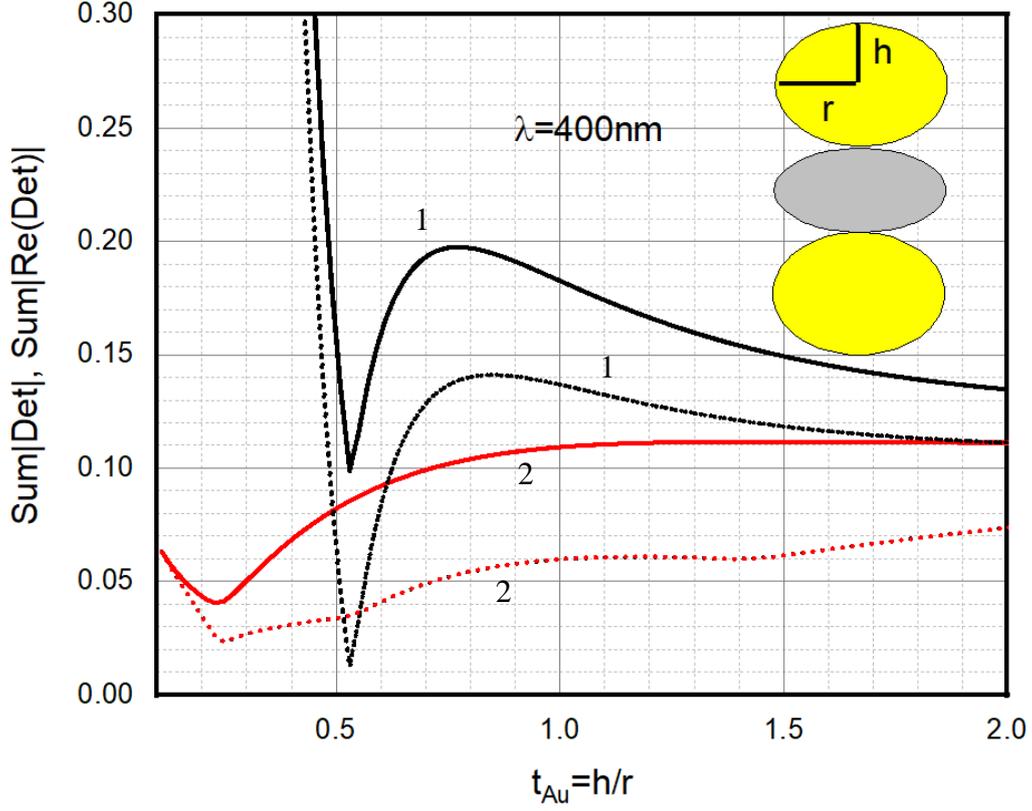

Fig. 3. Behavior of an objective function of the plasmonic structure, when the spacer nanoparticle is an ellipsoid of rotation, and up and down nanoparticles are gold nanoparticles of radius $r = 10$ nm. For black curves (1) eccentricity of the spacer is 0.25, and for red curves (2) – 2.0. For solid curves the objective function is $F_{compl}(\omega)$ (Eq.(31)), and for dashed curves – $F_{compl}^{(r)}(\omega)$ from Eq.(30).

If the structure consists of nanoparticles submonolayer cover of the magnetic film, the BLS signal can be calculated via the effective susceptibility of the cover $\mathrm{X}_{jk}^{(SC)}(k,z',\bar{\omega})$ as:

$$E_i^{BLS}(k,z_d,\bar{\omega}) \propto \int_{h_{min}}^{h_{max}} dz' G_{ij}^{(33)}(k,z,z',\bar{\omega}) \mathrm{X}_{jk}^{(SC)}(k,z',\bar{\omega}) E_k(k,z',\bar{\omega}) \,, \quad k = q_\parallel, \quad (36)$$

where $E_k(k,z',\bar{\omega})$ is a local field acting on the nanoplasmonic structure (submonolayer cover by nanoparticles) at the surface. As discussed above, this field consists of two components, namely the direct field of BLS:

$$E_i^{(BLS-0)}(k,z,\bar{\omega}) = -\bar{k}_0^2 \int_{h_f} d\mathbf{R}' G_{ij}^{(32)}(k,z,z',\bar{\omega}) \chi_{jl} E_l^{(0)}(k,z',\omega) \quad (37)$$

and the field induced by the dipole (Eq. (2)), which is enhanced by the nanoplasmonic structure field at fundamental frequency ω.

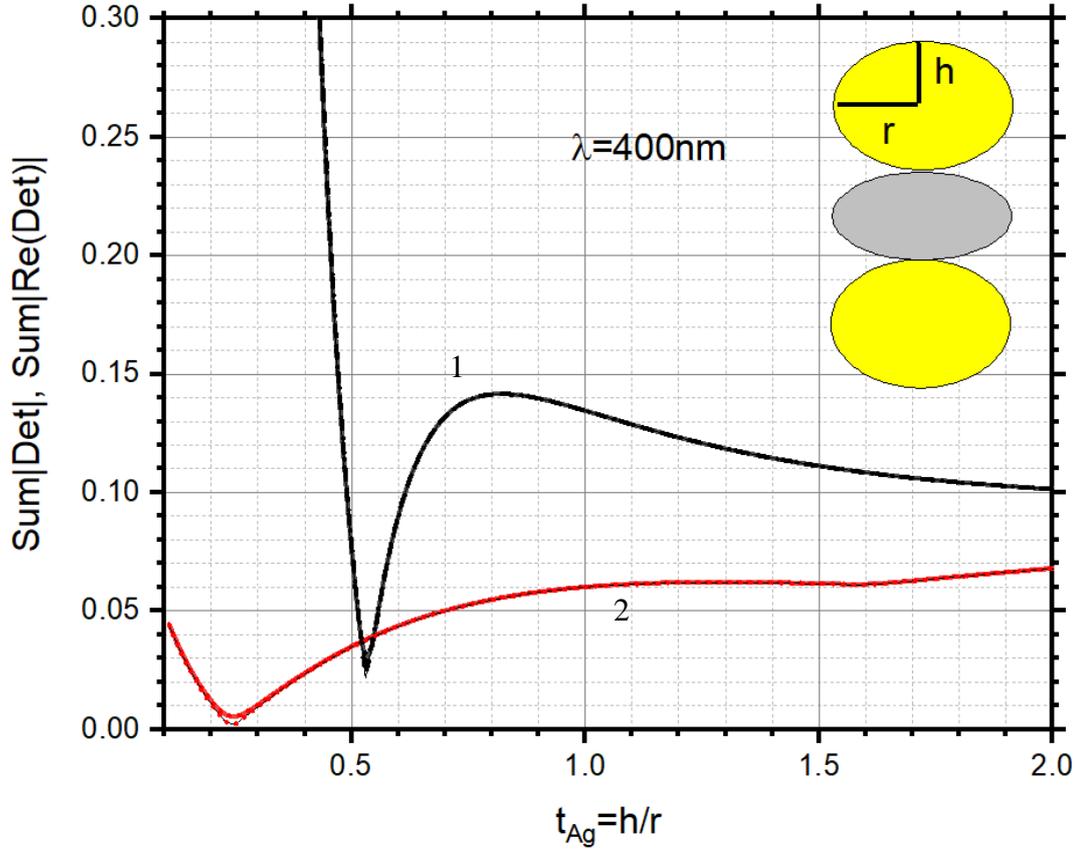

Fig. 4. Behavior of an objective function of the plasmonic structure, when the spacer nanoparticle is an ellipsoid of rotation, and up and down nanoparticles are silver nanoparticles of radius $r = 10$ nm. For black curves (1) eccentricity of the spacer is 0.25, and for red curves (2) – 2.0. For solid curves the objective function is $F_{compl}(\omega)$ (Eq.(31)), and for dashed curves – $F^{(r)}_{compl}(\omega)$ from Eq.(30).

Let us consider the formation of this field in more detail. The external field induces the density of a dipole moment in the nanoparticles submonolayer cover:

$$P_i^{(SC)}(k,z,\omega) = \varepsilon_0 X_{il}^{(SC)}(k,z',\omega)E_l^{(0)}(k,z',\omega) \; , \qquad (38)$$

where $X_{il}^{(SC)}(k,z',\omega)$ is the effective susceptibility of the submonolayer. This dipole momentum induces the electric field inside the magnetic film:

$$E_i^{(inc)}(k,\omega,z) = -i\omega\mu_0 \int_h dz' G_{ij}^{(23)}(k,z,z',\omega)(-i\omega)P_j^{(SC)}(k,z',\omega) =$$

$$= -k_0^2 \int_h dz' G_{ij}^{(23)}(k,z,z',\omega)X_{jl}^{(SC)}(k,z',\omega)E_l^{(0)}(k,z',\omega) \; . \qquad (39)$$

Integration in Eq. (39) is over the submonolayer of nanoparticles 'thickness'. Further, this field induces the dipole momentum at the shifted frequency (Eq. (2)):

$$P_i(\bar{\omega}) = -k_0^2 \frac{1}{4\pi} \chi_{ij} \int_h dz' G_{ij}^{(23)}(k,z,z',\omega) X_{jl}^{(SC)}(k,z',\omega) E_l^{(0)}(k,z',\omega). \quad (40)$$

Then, the oscillating dipole excites the BLS field at the detector, so that:

$$E_i^{(BLS-P)}(k,z,\bar{\omega}) = -\frac{k_0^2 \bar{k}_0^2}{4\pi} \int_{h_f} dz' G_{ij}^{(32)}(k,z,z',\bar{\omega}) \chi_{jk} \times$$

$$\times \int_h dz'' G_{km}^{(23)}(k,z',z'',\omega) X_{ml}^{(S)}(k,z'',\omega) E_l^{(0)}(k,z'',\omega). \quad (41)$$

The integration in the outer integral is over the thickness of the magnetic film. It should be emphasized that the field occurs both at the fundamental frequency (internal integral) and at the shifted frequency (outer integral). Thus, one can expect that the enhancement of BLS signal can be realized when the condition:

$$F^{(SC)}(k,\omega_\alpha) = \min\left[\left|D^{(SC)}(k,\omega_\alpha)\right|\right], \quad \omega_\alpha = \bar{\omega}, \text{ and/or } \omega_\alpha = \omega, \quad (42)$$

with

$$D^{(SC)}(k,\omega_\alpha) = \det\left[\left(\tilde{\chi}_{ii\atop zz}(\omega_\alpha)\right)^{-1} \delta_{ji\atop zz} + nk_0^2 G_{ji}^{(k)}(k,d,\omega_\alpha)\right] \quad (43)$$

is fulfilled.

Evidently, $\operatorname{Re} D^{(SC)}(k,\omega_i) = 0$ is a dispersion relation for the 'surface wave' located at the covered by the nanoparticles surface of a magnetic film. The Green's function $G_{ij}^{(\alpha\beta)}(k,z,z',\bar{\omega})$, when α or/and β equals 3, depends on the coordinate $z$ as $\exp\left[i\left(\sqrt{\bar{k}_0^2 - \left(k_\parallel^{BLS}\right)^2}\right)z\right]$, where $k_\parallel^{BLS}$ is an in-plane component of the scattered light, which depends on the incident angle ϑ. This indicates that in the context of the nanoplasmonic structure under consideration, the values of incident angles must be constrained by a requirement $\bar{k}_0^2 - (k_\parallel^{BLS})^2 \geq 0$. In other cases, the wave of the signal will transform into an evanescent wave (proportional to the $\exp\left[-\left(\sqrt{(k_\parallel^{BLS})^2 - \bar{k}_0^2}\right)z\right]$) and will not be registered by the detector. But if the measurements are performed in back-scattering mode, the vector $k_{\parallel}^{BLS} = -k_0 \sin \vartheta$ ($\vartheta$ is the angle of the incident light), $\bar{k}_0^2 - (k_{\parallel}^{BLS})^2 \geq 0$ and scattered light will be detected.

In conclusion, we derived the main equations for studying the conditions of the BLS signal enhancement, which will be performed in Part II of the work numerically.

**Discussion and conclusions**

In Part I of the work, we have developed a general approach to the theory of plasmon-enhanced Brillouin light scattering from a magnetic film. The method of the electrodynamic Green functions in the frame of effective susceptibility concept was used. The approach was focused on the acquisition and analysis of the effective susceptibility of the plasmon structures, which are considered as the elements of the BLS enhancement procedure. Analytical expressions for the optimization of the morphology of nanoplasmonic structures cover of the magnetic film have also been obtained.

The central concept of the work is that under conditions of localized (for single plasmonic nanostructures) or surface (for sub-monolayer coating with nanoparticles) plasmon resonance, effective enhancement of the BLS signal can be obtained due to the redistribution of the local field amplitude. However, with the implementation of the additional enhancement mechanisms (external illumination which forms a surface wave, or the utilization of one-dimensional photonic crystal on the surface, etc.), one can hope for a significant enhancement of the BLS by the nanoparticles covering the ferromagnetic film.

The numerical calculations of the optimization problem will be provided in the second part of the work [22]. That is, the optimization of the nanoplasmonic systems parameters range (size, shape, and mutual arrangement of parts of the structure, concentration of particles, etc.) allows to construct the most optimal structures to enhance the BLS signal. In Part II, based on the currently obtained results, we will elaborate on how to calculate the efficiency of the BLS signal enhancement by various nanoplasmonic structures, as described in Part I.

## Acknowledgments

V.L. thanks the Austrian Academy of Sciences' Joint Excellence in Science and Humanities (JESH), the ESI Special Research Fellowship for Ukrainian Scientists, and the IEEE Magnetics Society "Magnetism for Ukraine 2023" program for the support of this work. A.V.C acknowledges the Austrian Science Fund FWF for the support by the project I-6568 "Paramagnonics". The authors thank all the brave defenders of Ukraine, who made possible the finalization of the publication.